\documentclass[twocolumn,showpacs,preprintnumbers,amsmath,amssymb,prl] {revtex4}
\usepackage{graphicx}
\usepackage{dcolumn}
\usepackage{bm}
\usepackage{eucal}
\begin{document}

\title{One-particle irreducible density matrix for the spin disordered infinite $U$ Hubbard
chain}

\author{Vadim~V.~Cheianov$^{1}$ and M.B.~Zvonarev$^{2}$}
\affiliation{$^1$NORDITA, Blegdamsvej 17, Copenhagen {\O},  DK 2100, Denmark
\\ $^2$\O rsted Laboratory, Niels Bohr Institute for APG, Universitetsparken 5, Copenhagen {\O}, DK 2100, Denmark
}

\date{\today}

\begin{abstract}
In this Letter we present a calculation of the one-particle
irreducible density matrix $\rho(x)$ for the one-dimensional (1D)
Hubbard model in the infinite $U$ limit. We consider the zero
temperature spin disordered regime, which is obtained by first
taking the limit $U\to \infty$ and then the limit $T\to 0.$ Using
the determinant representation for $\rho(x)$ we derive analytical
expressions for both large and small $x$ at an arbitrary filling
factor $0<\varrho<1/2.$ The large $x$ asymptotics of $\rho(x)$ is
found to be remarkably accurate starting from
$x\sin(2\pi\varrho)\sim 1.$ We find that the one-particle momentum
distribution function $\rho(k)$ is a smooth function of $k$ peaked
at $k=2k_F,$ thus violating the Luttinger theorem.
\end{abstract}
\pacs{ 71.10.-w, 71.27.+a}
\maketitle

Recently, we reported results on the one-particle correlation
functions of the continuous 1D system of impenetrable spin $1/2$
fermions in the spin disordered regime
\cite{We-03short,We-03long}. It was found that the infrared
asymptotic behavior of the correlation functions, although
consistent with the assumption of spin-charge separation, is not
adequately described by the Luttinger model. This is to be
contrasted with the asymptotic behavior of the previously studied
correlation functions of the infinite $U$ Hubbard model in the
``antiferromagnetic'' ground state, understood as a limit of the
ground state of the Hubbard model as $U\to\infty$. In the latter
case the Luttinger model gives correct predictions
\cite{Ogata,Shiba,Korepin}.

In this paper we explore  the 1D Hubbard model \cite{Hubbard}
\begin{equation}
H=-\sum_{x;\sigma} \left(\psi^\dagger_{x,\sigma}\psi_{x+1,\sigma}+\mathrm{h.c.}\right) +
\sum_{x} U n_{x,\uparrow}n_{x,\downarrow},
\label{hamiltonian}
\end{equation}
in the limit $U\to\infty.$ Here $\psi_{x,\sigma}$ are fermion
fields with the spin index $\sigma=\uparrow, \downarrow,$ and
$n_{x,\sigma}= \psi_{x,\sigma}^{\dagger}\psi_{x,\sigma}$ are the
local fermion number operators. We will concentrate on the one
particle irreducible density matrix
\begin{equation}
\rho(x)=\langle \psi_{x,\uparrow}^{\dagger} \psi_{0,\uparrow}^{\phantom\dagger}\rangle,
\label{rhodef}
\end{equation}
at an arbitrary fixed filling factor
\begin{equation}
\varrho=\frac12\langle n_{x,\uparrow}+ n_{x,\downarrow}\rangle
\end{equation}
and in the limit $T\to 0.$

The ground state of the model \eqref{hamiltonian} at infinite $U$
is infinitely degenerate with respect to local spin rotations
\cite{Ogata}. Since the limit $T \to 0$ is taken after the limit
$U\to \infty,$ the thermal average $\langle\rangle$ in
\eqref{rhodef} reduces to the average over the infinitely
degenerate ground state. This is what we call an average taken in
the zero temperature spin disordered regime of the model.

Recently, the determinant representation for the dynamical
correlation functions of the infinite $U$ Hubbard model
\eqref{hamiltonian} in the spin disordered regime was obtained
\cite{IPA-98}. For the equal time correlation function
\eqref{rhodef} the determinant representation, given in
Ref~\cite{IPA-98}, can be written in the following form:
\begin{equation}
\rho(x)=\frac{1}{8\pi i} \oint_{|z|=1} \frac { d z}
{z} F(z) B_{--}(z)\det(\hat I+\hat V)(z). \label{rhointz}
\end{equation}
Here the function $F(z)$ is
\begin{equation}
\label{Fdef} F(z)=1+\frac{z}{2-z}+\frac{1}{2 z-1}.
\end{equation}
The determinant
\begin{eqnarray}
\det (\hat I +\hat V) = \sum_{N=0}^{\infty}\frac1{N!}
\int_{-K}^{K}{\rm d} k_1\ldots\int_{-K}^{K}{\rm d} k_N \nonumber\\
\times\det\left[
\begin{matrix}
V(k_1,k_1)& \cdots & V(k_1,k_N)\cr \vdots & \ddots & \vdots \cr
V(k_N,k_1)& \cdots & V(k_N,k_N)
\end{matrix}
\right] \label{det}
\end{eqnarray}
is the Fredholm determinant of a linear integral operator $\hat V$
with the kernel
\begin{equation}
V(k,p)=\frac{e_+(k)e_-(p)-e_+(p)e_-(k)}{2\tan\left[\frac{1}{2}(k-p)\right]}
\label{Vdef}
\end{equation}
defined on $[-K,K]\times[-K, K].$ Here
\begin{equation}
K=2\pi \varrho
\end{equation}
is twice the Fermi momentum. The functions $e_{\pm}$ entering
Eq.~\eqref{Vdef} are defined as follows
\begin{align}
& e_-(k)=\frac{1}{\sqrt \pi} e^{-i k x/2}, \label{emdef}
\\
& e_+(k)=\frac{i}{2}\frac{1}{\sqrt \pi} e^{i k x /2} (1-z).
\label{epdef}
\end{align}
The function $B_{--}(z)$ is
\begin{equation}
B_{--}(z) = \int_{-K}^{K} {\rm d} k e_{-}(k) (\hat I + \hat
V)^{-1} e_{-}(k) \label{Bmmdef}
\end{equation}
Consider the contour integral in Eq. \eqref{rhointz}. According to
definitions Eqs. \eqref{Vdef} and \eqref{epdef} the Fredholm
operator $\hat V$ is linear in $z.$ This implies \cite{We-03long}
that the product $B_{--}(z)\det(\hat I- \hat V) (z) $ is analytic
in the complex  $z$-plane. Therefore, the integral is given by the
residue of the integrand at the pole $z=1/2$ of the function
$F(z)$
\begin{equation}
\rho(x)=\frac{1}{4}B_{--}(1/2)\det(\hat I+\hat V)(1/2).
\label{residue}
\end{equation}

Consider the {\it short distance} behavior of $\rho(x)$ first. For
any $x$ the kernel \eqref{Vdef} can be written as a sum of $2 x$
separable kernels (recall that $x$ is a discrete variable,
$x=0,1,2,\ldots$)
\begin{equation}
V(k,p)=\frac{z-1}{4 \pi}
\sum_{m=1}^{2x} u_m(k) u^{*}_m(p),
\end{equation}
where
\begin{equation}
u_m(k)=\left\{
 \begin{array}{cl}
 e^{i(m-\frac{x}{2})k}, &\quad m=1,\ldots, x \\
 e^{-i(m-\frac{3x}{2})}, &\quad m=x+1,\ldots,2x
 \end{array}.
\right.
\end{equation}
Therefore, $\det(\hat I+\hat V) $ can be expressed in terms of the
determinant
\begin{equation}
\det(\hat I+\hat V)= \det\nolimits_{2x}(\mathbf I +\mathbf V)
\label{detfrombfV}
\end{equation}
of an $2x \times 2x$ matrix $\mathbf V$:
\begin{equation}
\mathbf V= \frac{z-1}{2 \pi} \left(\begin{array}{c|c} Q&P \\
\hline
P&Q
\end{array}\right).
\label{bfVdef}
\end{equation}
Here $Q$ and $P$ are $x\times x$ matrices with the entries defined
by
\begin{align}
Q_{mn}&=\frac{\sin[K(m-n)]}{m-n}, &\quad n,m=1,\ldots,x
\label{Qdef}
\\
P_{mn}&=\frac{\sin[K(m+n-x)]}{m+n-x}, &\quad n,m=1,\ldots,x
\label{Pdef}
\end{align}
where
\begin{equation}
Q_{nn}=P_{(x-n)n}=K.
\end{equation}
For $B_{--}(z)$ one has
\begin{equation}
B_{--}=\frac{2 \sin Kx}{\pi x}- \frac{z-1}{4 \pi} \mathbf
a^{\mathrm T}  (\mathbf I+\mathbf V)^{-1} \mathbf b,
\label{BmmfrombfV}
\end{equation}
where the $2x$-dimensional vectors $\mathbf a$ and $\mathbf b$ are
defined by
\begin{equation}
\mathbf a_n= \left \{
\begin{array}{cl}\displaystyle{
\frac{2 \sin K n}{\sqrt{\pi} n}}, &\quad n=1, \ldots, x
\\
\\
\displaystyle{ \frac{2 \sin K(n-2x)}{\sqrt{\pi}(n-2x)}}, &\quad
n=x+1, \ldots , 2x
\end{array}
\right.
\label{adef}
\end{equation}
and
\begin{equation}
\mathbf b_n=\frac{2 \sin K (n-x)}{\sqrt{\pi} (n-x)}, \quad n=1,
\dots,2 x. \label{bdef}
\end{equation}
Eqs.~\eqref{detfrombfV} through \eqref{bdef} combined with
\eqref{residue} are convenient for the calculation of $\rho(x)$ at
small enough $x.$ For example,
\begin{align}
&\rho(0)=\frac{K}{2 \pi}, \\
&\rho(1)=\frac{\sin K}{2\pi}, \\
&\rho(2)=\frac{\sin^2 K}{4 \pi^2}+\frac{(2\pi -K)\sin2K}{8 \pi^2}.
\end{align}
With increasing $x$ the complexity of the exact expression for
$\rho(x)$ grows rapidly.

Next, we calculate the {\it long distance} asymptotics of the
density matrix \eqref{rhodef} using the determinant representation
\eqref{rhointz}. Technically, the asymptotic analysis will be
similar to that carried out for the continuous limit of the model
in Ref.~\cite{We-03long}.

To calculate $\det(\hat I+\hat V)$ write the difference equation
for the kernel (\ref{Vdef}):
\begin{align}
V(k,p;x+1)=&e^{\frac{i}{2}(k-p)} V(k,p;x)
\nonumber
\\
&+ i e_{-}(k;x)e_{+}(p;x) \cos\frac{k-p}{2}. \label{kerneleq}
\end{align}
From this equation it follows that
\begin{equation}
{\det(\hat I+\hat V)(x+1;z)} = {\det(\hat I+\hat V)(x;z)} W(x;z),
\label{deteq}
\end{equation}
where
\begin{equation}
W(x)=
\det\left[
\begin{array}{cc}
1+\frac{i}{2} B_{+-}(x) & \frac{i}{2} D_{-+}(x) \\
\frac{i}{2} C_{+-}(x) &  1+\frac{i}{2} A_{-+}(x)
\end{array}
\right]
\label{Wdef}
\end{equation}
and
\begin{align}
A_{ab}&=\int_{-K}^{K} d k e_a(k)e^{-i k} (\hat I+\hat V)^{-1}[e^{ik}e_b(k)] \label{A},\\
B_{ab}&=\int_{-K}^{K} d k e_a(k) (\hat I+\hat V)^{-1}e_b(k) \label{B},\\
C_{ab}&=\int_{-K}^{K} d k e_a(k)e^{i k} (\hat I+\hat V)^{-1}e_b(k) \label{C},\\
D_{ab}&=\int_{-K}^{K} d k e_a(k)e^{-i k} (\hat I+\hat
V)^{-1}e_b(k). \label{D}
\end{align}
The indices $a$ and $b$ run through two values: $a,b=\pm.$

\begin{figure}
\includegraphics[width=0.45\textwidth]{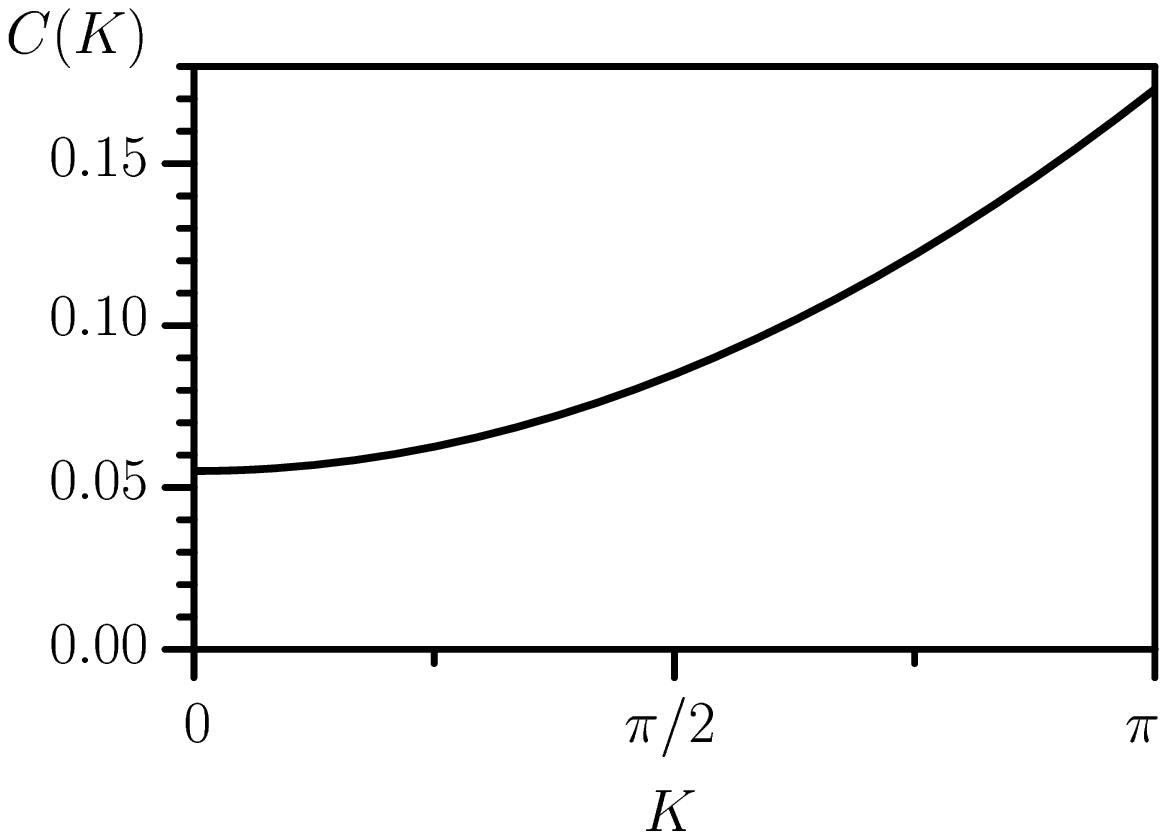}
\caption{\label{fig:Constant} The constant $C(K)$ in Eqs.
\eqref{detansw} and \eqref{bansw} plotted as a function of $K.$
The plot data were obtained from the comparison of the asymptotic
formula \eqref{detansw} with the numerically calculated Fredholm
determinant \eqref{det}.}
\end{figure}

The resolvent operator $(\hat I+\hat V)^{-1}$ and, therefore, the
functions (\ref{A})-(\ref{D}) can be found from the solution of
the corresponding matrix Riemann-Hilbert problem \cite{KBI}. The
scheme of the asymptotic solution of the matrix Riemann-Hilbert
problem associated with the kernel \eqref{Vdef} is very similar to
the one given in \cite{We-03long}. It is based on the non-linear
steepest-descend method \cite{Deift}. The main results of the
asymptotic analysis are as follows. For $z=1/2$
\begin{equation}
W(x;z=1/2)=2^{-{K}/{\pi}}\left[1+ \frac{\nu^2}{2 x}\right]+\delta W(x),
\label{W}
\end{equation}
where
\begin{equation}
\nu=\frac{\ln 2}{\pi}.
\end{equation}
The error term  $\delta W(x)$ decays as $x^{-2}$ for $ x\sin K
>1.$ Solving Eq.~\eqref{deteq} with $W$ given by Eq.~\eqref{W} one
gets in the large $x$ limit
\begin{equation}
\det(\hat I+\hat V)(x)=e^{C(K)} (\sin K)^{\frac{\nu^2}{2}} \cdot
2^{-\frac{K }{\pi} x} x^{\frac{\nu^2}{2}}, \label{detansw}
\end{equation}
where $C(K)$ is independent of $x.$  Numerically, $\exp[C(K)]$ is
close to unity for all $K,$ as can be seen in
Fig.~\ref{fig:Constant}.
The exact expression for $C(0)$ is given in Ref.~\cite{We-03long}
and is, numerically, equal to $0.0550839\ldots.$ This agrees
perfectly with Fig.~\ref{fig:Constant}.

The asymptotic formula for the one particle density matrix
\eqref{rhointz} reads
\begin{align}
\rho(x)&=\frac{i \pi 2 \sqrt 2 e^{C(K)}(\sin K)^{\frac{\nu^2}{2}}}
{\cosh^2 \left( K \nu /2 \right)} e^{-\nu K x} x^{\frac{\nu^2}{2}}
\nonumber
\\
\times&
\left[
\frac{ (2 \sin K)^{-i \nu}}{ \Gamma(-i \nu /2)^2}
\frac{e^{i K x}}{x^{1+i \nu}} -
\frac{ (2 \sin K)^{i \nu}}{ \Gamma(i \nu /2)^2}
\frac{e^{-i K x}}{x^{1-i \nu}}
\right]
\label{bansw}
\end{align}
with the relative correction of the order of $x^{-1}.$ The formula
\eqref{bansw} is the main result of the paper.

Let us discuss Eq.~\eqref{bansw}. The structure of the correlation
function is essentially the same as for the impenetrable fermion
gas \cite{Berkovich,We-03short,We-03long}. The correlation
function contains the exponentially decaying factor $\exp(-\nu K
x),$ and factors obeying the power law scaling. The complex-valued
anomalous exponents do not depend on $K$ or, equivalently, on the
filling factor $\varrho.$ A similar situation takes place for the
infinite $U$ Hubbard model in the Luttinger regime: the Luttinger
scaling exponents do not depend on the filling factor
\cite{Korepin}. The results for the continuous model, impenetrable
fermion gas \cite{Berkovich,We-03short,We-03long}, can be
recovered by taking the limit $K\to0$ in Eq.~\eqref{bansw} at a
fixed $Kx.$

\begin{figure}
\includegraphics[width=0.45\textwidth]{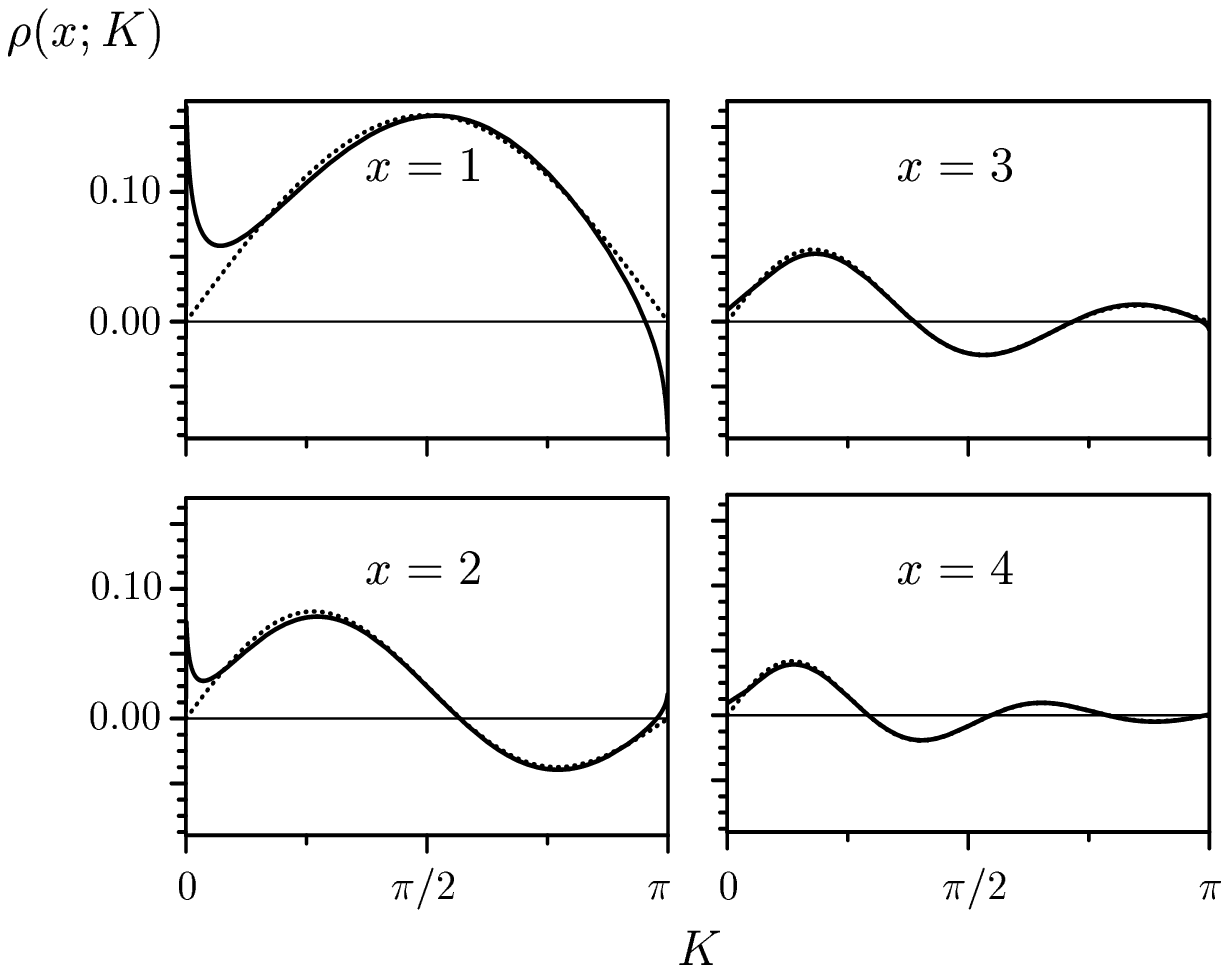}
\caption{\label{fig2}  Density matrix $\rho(x)$ plotted as a
function of $K$ for $x=1,\dots, 4.$ The  asymptotic result
\eqref{bansw} (solid line) is in good agreement with the exact
result (dotted line) even for small $x\sin K.$}
\end{figure}

Formally, the asymptotic formula given in Eq.~\eqref{bansw} is
valid for $x\sin K\gg1.$ Nevertheless, it is remarkably good even
for $x\sin K\sim1$ as can be seen from Fig.~\ref{fig2},
where the exact expression obtained from Eqs. \eqref{detfrombfV}
through \eqref{bdef} is compared with the asymptotics
Eq.~\eqref{bansw}.

\begin{figure}
\includegraphics[width=0.45\textwidth]{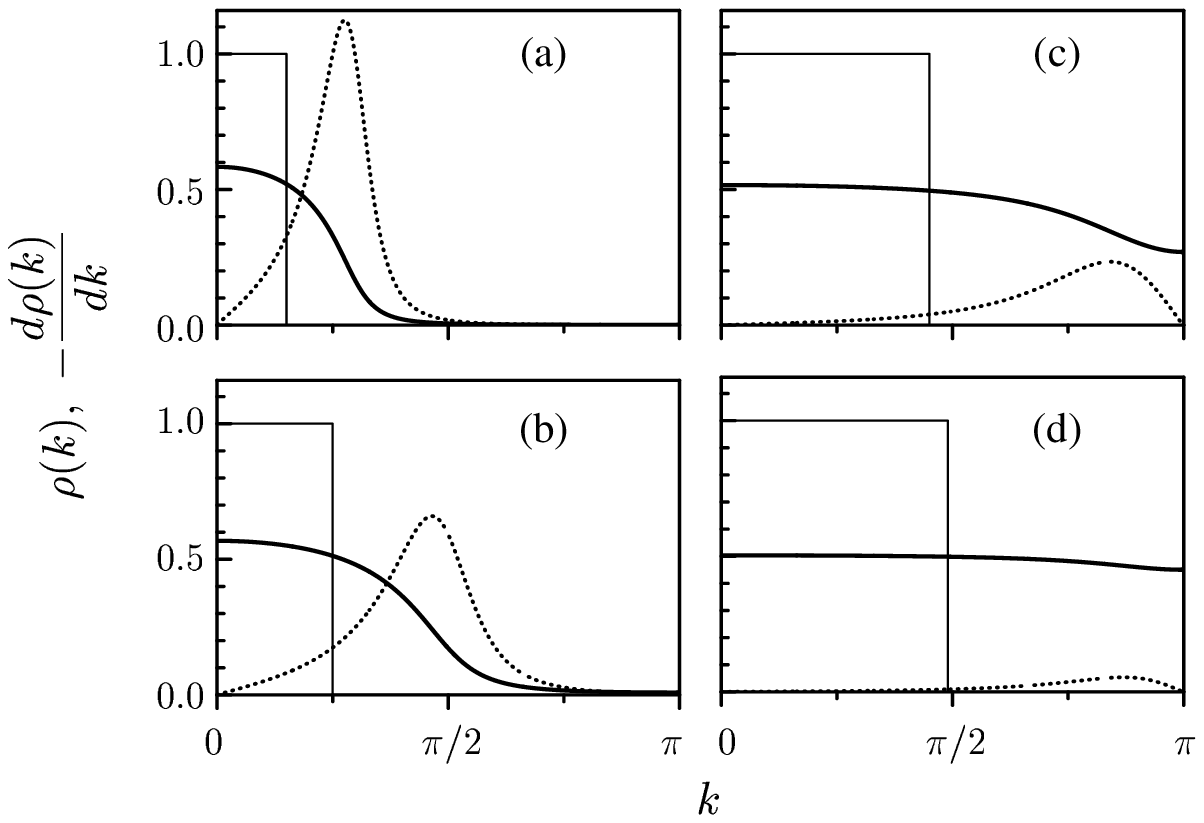}
\caption{\label{fig3} Momentum distribution function $\rho(k)$
(thick line) and its derivative $-d\rho(k)/dk$ (dotted line) for
different filling factors $\varrho:$ (a) $\varrho=1/8$ (b)
$\varrho=1/4$ (c) $\varrho=0.45$ (d) $\varrho=0.49$. The
Fermi-Dirac distribution (thin line) corresponding to these
filling factors is shown for comparison. The function $\rho(k)$
satisfies $\rho(k)=\rho(-k).$}
\end{figure}

Finally, consider the momentum distribution function
\begin{equation}
\rho(k)=\sum_{x=-\infty}^{\infty} e^{-ikx} \rho(x).
\end{equation}
Due to the exponentially decaying term in the asymptotic
expression Eq.~\eqref{bansw}, the function $\rho(k)$ is continuous
with all its derivatives for all $k.$ Combining the short distance
representation Eqs.~\eqref{detfrombfV}-\eqref{bdef} and the long
distance expansion \eqref{bansw} we plot $\rho(k)$ for $0\le
k\le\pi$ at different filling factors $\varrho$ in
Fig.~\ref{fig3}.
Note that the smoothness of $\rho(k)$ in the spin disordered
regime, considered here, is in contrast with the Luttinger regime
considered in Ref~\cite{Ogata}, where $d\rho(k)/dk$ is singular at
the Fermi momentum $k_F=K/2,$ in accordance with the Luttinger
theorem \cite{Blagoev}. Another peculiarity of the spin disordered
regime is that $d\rho(k)/dk$ is peaked around $k=2 k_F$ as it can
be seen in Fig.~\ref{fig3}. This can be viewed as a mild violation
of the Luttinger theorem for this system.

We thank A. Luther for helpful discussions. M.B. Zvonarev's work
was supported by the Danish Technical Research Council via the
Framework Programme on Superconductivity.


\begin{thebibliography}{99}
\bibitem{We-03short} V.V. Cheianov and M.B. Zvonarev, e-print: cond-mat/0308470.
\bibitem{We-03long} V.V. Cheianov and M.B. Zvonarev, e-print: cond-mat/0310499.
\bibitem{Ogata} M. Ogata and H. Shiba, Phys. Rev. B {\bf 41}, 2326 (1990).
\bibitem{Shiba} K. Penc, F. Mila, and H. Shiba, Phys. Rev. Lett. {\bf 75}, 894 (1995);
K. Penc, {\it et al.}, Phys. Rev. B {\bf 55}, 15475 (1997).
\bibitem{Korepin} H. Frahm and V.E. Korepin, Phys. Rev. B, {\bf  42}, 10553 (1990);
N. Kawakami and S.K. Yang, Phys. Lett. A {\bf 148}, 359 (1990);
H.J. Schulz, Phys. Rev. Lett. {\bf 64}, 2831 (1990).
\bibitem{Hubbard}   J. Hubbard, Proc. R. Soc. London A {\bf 276}, 238(1963)
\bibitem{IPA-98} A.G. Izergin, A.G. Pronko, and N.I. Abarenkova,
Phys. Lett. A {\bf 245}, 537 (1998).
\bibitem{KBI} V.E. Korepin, N.M. Bogoliubov and A.G. Izergin, {\it
Quantum Inverse Scattering Method and Correlation Functions}
(Cambridge University Press, Cambridge, England, 1993).
\bibitem{Deift} P. Deift, T. Kriecherbauer, K. T-R McLaughlin, S. Venakides, and X. Zhou,
Commun. Pure Appl. Math. {\bf 52}, 1491 (1999); P. Deift, {\it
Orthogonal Polynomials and Random Matrices:A Riemann-Hilbert
Approach} (Courant Institute of Mathematical Sciences, New York,
NY; American Mathematical Society, Providence, RI, 1999).
\bibitem{Berkovich} A. Berkovich and J.H. Lowenstein,
Nucl. Phys. B {\bf 285}, 70 (1987); A. Berkovich, J. Phys. A:
Math. Gen. {\bf 24}, 1543 (1991).
\bibitem{Blagoev}K.B. Blagoev and K.S. Bedell, Phys. Rev. Lett. {\bf 79}, 1106 (1997).
\end{thebibliography}
\end{document}